\documentclass[journal,twoside,web]{ieeecolor}
\usepackage{amsmath,amssymb,amsfonts}
\usepackage{algorithmic}
\usepackage{graphicx}
\usepackage{textcomp}
\usepackage{booktabs}
\usepackage{adjustbox}
\usepackage{dblfloatfix}
\usepackage{comment}
\usepackage{capt-of}
\usepackage{threeparttable}
\usepackage[utf8]{inputenc}
\usepackage{csquotes}
\usepackage[english]{babel}
\usepackage[backend=biber, style=ieee, url=false, doi=false, maxbibnames=6]{biblatex}

\def\logoname{LOGO-generic-web}
\definecolor{subsectioncolor}{rgb}{0,0.541,0.855}
\def\journalname{Generic Colorized Journal}

\def\BibTeX{{\rm B\kern-.05em{\sc i\kern-.025em b}\kern-.08em
    T\kern-.1667em\lower.7ex\hbox{E}\kern-.125emX}}
\begin{document}
\title{FundusQ-Net: a Regression Quality Assessment Deep Learning Algorithm for Fundus Images Quality Grading}
\author{Or Abramovich, Hadas Pizem, Jan Van Eijgen, \textcolor{black}{Ilan Oren, Joshua Melamed}, Ingeborg Stalmans, Eytan Z. Blumenthal and Joachim A. Behar
\thanks{Submitted for review on January 25th, 2023}
\thanks{Joachim A. Behar is with the Technion – Israel Institute of Technology, Haifa, Israel (e-mail: jbehar@technion.ac.il).}}

\maketitle

\begin{abstract}
Objective: Ophthalmological pathologies such as glaucoma, diabetic retinopathy and age-related macular degeneration are major causes of blindness and vision impairment. There is a need for novel decision support tools that can simplify and speed up the diagnosis of these pathologies. A key step in this process is to automatically estimate the quality of the fundus images to make sure these are interpretable by a human operator or a machine learning model. We present a novel fundus image quality scale and deep learning (DL) model that can estimate fundus image quality relative to this new scale. 

Methods: A total of 1,245 images were graded for quality by two ophthalmologists within the range 1-10, with a resolution of 0.5. A DL regression model was trained for fundus image quality assessment. The architecture used was Inception-V3. The model was developed using a total of 89,947 images from 6 databases, of which 1,245 were labeled by the specialists and the remaining 88,702 images were used for pre-training and semi-supervised learning. The final DL model was evaluated on an internal test set (n=209) as well as an external test set (n=194).  

Results: The final DL model, denoted FundusQ-Net, achieved a mean absolute error of 0.61 (0.54-0.68) on the internal test set. When evaluated as a binary classification model on the public DRIMDB database as an external test set the model obtained an accuracy of 99\%. 

Significance: the proposed algorithm provides a new robust tool for automated quality grading of fundus images. 
\end{abstract}

\begin{IEEEkeywords}
Fundus image, quality assessment, deep learning, semi supervised learning.
\end{IEEEkeywords}

\section{Introduction}
\label{sec:introduction}
\IEEEPARstart{O}{cular} pathologies are a leading cause of visual impairment and blindness globally, with 237.1 million people suffering from moderate or severe visual impairment and 38.5 million people blind  \cite{Flaxman2017GlobalMeta-analysis}. Age-related macular degeneration (AMD), cataract, diabetic retinopathy, and glaucoma are among the pathologies with the most severe impact on visual acuity \cite{Flaxman2017GlobalMeta-analysis}. A variety of techniques are used for diagnosis, including optical coherence tomography (OCT) imaging, color fundus photography, fluorescein angiography (FA), fundus autofluorescence (FAF) and optical coherence tomography angiography (OCTA) \cite{Salz2015ImagingRetinopathy} \cite{Spaide2003FundusDegeneration} \cite{deCarlo2015AOCTA}. Among these methods, the computerized analysis of color fundus images is a widely used method for diagnosis \cite{Kaur2021DiabeticReview} \cite{Guven2013AutomaticImages} \cite{Raghavendra2018DeepImages}.
A digital fundus image (DFI) is an image of the inner lining of the eye which captures the optic disc, the fovea, the macula, the retina and blood vessels. Early detection and treatment  of ocular pathologies can help slow down and sometimes prevent further vision loss \cite{Stein2021GlaucomaReview}. However, there is currently a global shortage of ophthalmologists which, according to current trends, will only intensify in the following years \cite{Resnikoff2012ThePractitioners}. This prevents many people from being diagnosed in a timely manner. To address this issue, techniques such as automated screening devices and telemedicine have been proposed as a way to provide fast diagnoses without the need for an on-site ophthalmologist \cite{Saleem2020VirtualEra}.
Real-world DFIs can be of low quality due to a variety of factors, including dirty camera lenses, improper flash and gamma adjustment, eye blinks, and occlusion by eyelashes \cite{Raj2019FundusScope}, as well as media opacity and insufficient technician skill \cite{Strauss2007ImagePhotography}. Therefore, for usage in a large-scale screening or telemedicine devices, the device must be able to automatically identify and handle low-quality images. In reality, state-of-the-art approaches do not currently meet this requirement. Researchers often manually discard low-quality DFIs from their databases as a preliminary step  \cite{Liu2019DevelopmentPhotographs} \cite{Li2018EfficacyPhotographs}. For example, Liu et al. \cite{Liu2019DevelopmentPhotographs} manually reviewed 274,413 DFIs for their research, employing several tiers of human graders, and discarded 1.7\% of them (n=4,812). Li et al. \cite{Li2018EfficacyPhotographs} trained 21 ophthalmologists to discern between gradable and ungradable DFIs. These ophthalmologists reviewed 48,116 DFIs. This is a time-consuming process that limits the clinical applicability of developed algorithms. 

\begin{figure*}[!t]
\centerline{\includegraphics[width=\textwidth]{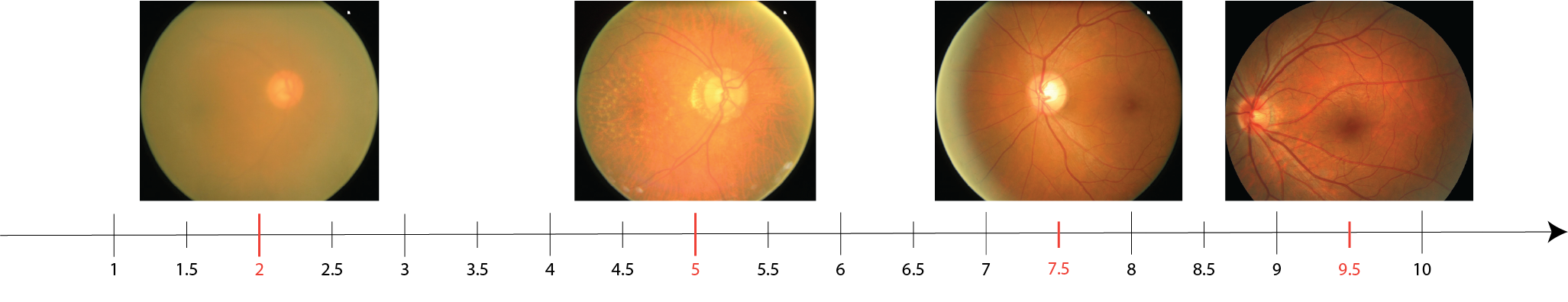}}
\caption{An example of different DFIs used for the reference set. Image A is graded as 2, image B is graded as 5, image C is graded as 7.5 and image D is graded as 9.5. \textcolor{black}{Images A, B and C are taken from the ORIGA database \cite{Zhang2010ORIGA-lightResearch}. Image D is taken from the REFUGE dataset \cite{Orlando2020REFUGEPhotographs}.}}
\label{quality_scale_ref}
\end{figure*}

\subsection{State-of-the-art}
Analyzing the quality of a DFI is a fundamentally different task than analyzing the quality of a regular image. Raj et al. \cite{Raj2019FundusScope} explained that typical Image Quality Assessment (IQA) methods might not be adequate for Retinal Image Quality Assessment (RIQA) because the statistical properties of DFIs differ vastly from that of natural images. This means that field specific methods must be developed.

RIQA algorithms can be separated into three groups: similarity-based, segmentation-based and ML-based. Similarity-based algorithms compare the target image to a set of high quality images. Segmentation-based algorithms first extract structures from the target image, and then analyse them according to different parameters. ML-based algorithms involve training an ML model on either extracted features from the DFI or the DFI itself. 

Similarity-based and segmentation-based methods are not commonly used nowadays \cite{Raj2019FundusScope}. Similarity-based algorithms are not popular because they fail to take into consideration structural information contained in the DFI. Segmentation-based algorithms are very rigid, and only function when the fundus has certain characteristics, such as a specific shape, size and location of physiological features within the image. Changes to these parameters leads to a reduction in the algorithm's performance \cite{Raj2019FundusScope}.

These factors, in conjunction with advancements in hardware and convolutional neural network (CNN) architectures, have led to the rise of ML-based algorithms, and specifically CNN-based algorithms \cite{Chan2021DeepDisorders}. In the following section we review state-of-the-art ML-based models used for RIQA.


In 2019, Fu et al. \cite{Fu2019EvaluationColor-spaces} used the novel MCF-Net architecture and the EyeQ database, a subset of EyePACS, which contains 28,792 DFIs, using a train/test split of 43.5/56.5. They experimented with using three different color spaces for the purpose of quality assessment: RGB, HSV and LAB. The MCF-Net is comprised of three parallel CNNs, each of which analyzes a DFI from a certain color space. The network then fuses the features from each color space together, and returns the result. The authors used three classes: ``Good", ``Usable" and ``Reject", and asked two experts to grade the quality of the DFIs. They then tested their network against several different architectures which only accept a single color space. The authors achieved an accuracy of 91.8\%.

In 2020, Zapata et al. \cite{Zapata2020ArtificialGlaucoma} used a novel CNN architecture called CNN-1 and the Optretina database, which contains 306,302 DFIs. 150,075 of them were labeled for quality, using the labels ``Good" and ``Bad". They reported an AUC of 0.947 and accuracy of 91.8\% using 10-fold cross validation. 

In 2021, Karlsson et al. \cite{Karlsson2021AutomaticScale} proposed a novel continuous quality scale. The quality score lay on a scale of 0.0 to 1.0, taking into consideration two features: focus and contrast. In their work they first extracted the relevant features using various filters, and then used a random forest regression algorithm to estimate the score for each feature. Their database consisted of 787 retinal oximetry images, which is an imaging modality closely related to fundus photography, and 253 DFIs. After choosing the threshold of 0.625, they measured their results on the binary-labeled DRIMDB database and achieved an accuracy of 98.1\%.

\textcolor{black}{In 2022, Shi et al. \cite{Shi2022AssessmentAnalysis} introduced the ARIA model. In addition to distinguishing between “Good” and “Bad” quality DFIs, the model was also capable of distinguishing between eye-abnormality-associated-poor-quality and artifact-associated-poor-quality on color fundus retinal images. The study utilized a database of 2434 retinal images, including 1439 good quality and 995 bad quality images. The method achieved a sensitivity, specificity, and accuracy of 98.0\%, 99.1\%, and 98.6\%, respectively, for distinguishing between good and poor quality images and 92.2\%, 93.8\%, and 93\%, respectively, for differentiating between the two types of poor quality images. In external validation, the ARIA model achieved an AUC of 0.997 for overall quality classification and 0.915 for the classification of the two types of poor quality.}

In 2023, Guo et al. \cite{Guo2023LearningRegularization} proposed a deep CNN based method DFI quality assessment. The method utilizes a dual-path CNN architecture with attention blocks, as well as label smoothing and cost-sensitive regularization techniques to improve performance. The authors used the EyeQ dataset, which has been previously used Fu et al. \cite{Fu2019EvaluationColor-spaces}, using the same train/test split. In addition, they have manually annotated 20,000 DFIs which were used for the train set. They reported a 0.91 Kappa score, and precision, recall, and F1-score of 0.89, 0.88, and 0.88 respectively.

To our knowledge there exist two open access databases for evaluating fundus quality assessment algorithms: DRIMDB and EyeQ \cite{Fu2019EvaluationColor-spaces} \cite{Sevik2014IdentificationMethods}. DRIMDB consists of 216 DFIs, which is insufficient for DL, and EyeQ was originally created as a diabetic retinopathy database, and uses a trinary labeling system: ``Good", ``Usable" and ``Bad". 

\begin{table*}[b!]
\centering
\begin{threeparttable}
\begin{adjustbox}{max width=\textwidth}
\begin{tabular}{@{}cccccccc@{}}
\toprule
Name & Num. Images & Num. Images Used & \textcolor{black}{FOV} & \textcolor{black}{Resolution} & Grading Method & Mean/SD Grades & Purpose \\ \midrule
Drishti-GS \cite{Sivaswamy2014Drishti-GS:Segmentation} & 101 & 21 & \textcolor{black}{30} & \textcolor{black}{2047x1760} & EZB Quality Scale & 6.60/1.11 & Training and test examples \\ \midrule
ORIGA \cite{Zhang2010ORIGA-lightResearch} & 650 & 145 & \textcolor{black}{45} & \textcolor{black}{2484x2048 (On Average)} & EZB Quality Scale & 7.43/0.93 & Training and test examples \\ \midrule
REFUGE \cite{Orlando2020REFUGEPhotographs} & 1,200 & 84 & \textcolor{black}{-} & \textcolor{black}{2124x2056} & EZB Quality Scale & 7.03/1.37 & Training and test exampls \\ \midrule
LEUVEN & 37,345 & 1,000 & \textcolor{black}{30} & \textcolor{black}{1444x1444} & EZB Quality Scale & 6.01/2.4 & Training and test examples \\ \midrule
EyeQ \cite{Fu2019EvaluationColor-spaces} & 28,792 & 28,792 & \textcolor{black}{-} & \textcolor{black}{3636x2473 (On Average)} & Good/Usable/Bad & - & Transfer learning examples \\ \midrule
EyePACS \cite{Cuadros2009EyePACS:Screening:} & 88,702 & 59,910 & \textcolor{black}{-} & \textcolor{black}{3636x2473 (On Average)} & Not Graded & - & Semi supervised learning examples \\ \midrule
DRIMDB \cite{Sevik2014IdentificationMethods} & 216 & 194 & \textcolor{black}{60} & \textcolor{black}{570x760} & Good/Bad & - & External test set examples \\ \bottomrule
\end{tabular}
\end{adjustbox}
\begin{tablenotes}
   \item[*] EZB Quality Scale is the scale introduced and used in this research.  
  \end{tablenotes}
\caption{Databases used in this research}
\label{tab:databases}
\end{threeparttable}
\end{table*}

\subsection{Limitations of previous work}
The assessment of fundus image quality has traditionally been approached using binary classification, which assigns images as either ``good" or ``bad" quality \cite{Raj2019FundusScope} \cite{Chan2021DeepDisorders}. However, this approach has several limitations. Firstly, the definition of ``good" and ``bad" quality is subjective and can vary depending on the pathology being studied and the individual ophthalmologist making the assessment. Secondly, this approach ignores the fact that image quality is a continuous variable, and as such binary labels may be applied to images of significantly different quality. This can result in errors when classifying images with borderline quality. In addition, a numerical assessment of image quality could potentially be used in other applications, such as confidence estimation in diagnostic classification tasks.

In contrast to previous work, such as Karlsson et al. \cite{Karlsson2021AutomaticScale}, our method considers a wider range of factors in the calculation of the quality score, including noise, uneven illumination, and artifacts. This enables for a more comprehensive assessment of image quality and may be important in clinical decision making, particularly with regards to the visibility of intra-ocular structures such as the macula and optic disc.

It can also be challenging to accurately benchmark the performance of algorithms in this field. Many researchers report their results using a private test set that has been labeled using their own unique standards. If external validation is sought, the DRIMDB dataset is often used due to its binary labeling and public availability \cite{Karlsson2021AutomaticScale} \cite{Sevik2014IdentificationMethods}  \cite{Shao2017AutomatedStructure} \cite{Chalakkal2019QualityLearning}.

In this study, our goal is to develop a generic and flexible data-driven algorithm for fundus image quality assessment. The quality scale used in this work is available to other researchers at the following repository: https://github.com/aim-lab/FundusQ-Net-Quality-Scale.

\section{Materials and Methods}

\subsection{New quality scale}
In our work, we employed an objective approach, which was built upon subjective principles. Utilizing the fundus photograph open databases (Drishti-GS, ORIGA and REFUGE), a total of 28 DFIs were selected to span a scale from 1 to 10 with 1 being the lowest possible quality score and 10 the highest. Increments of 0.5 were considered.  

\begin{figure}
\centerline{\includegraphics[width=\columnwidth]{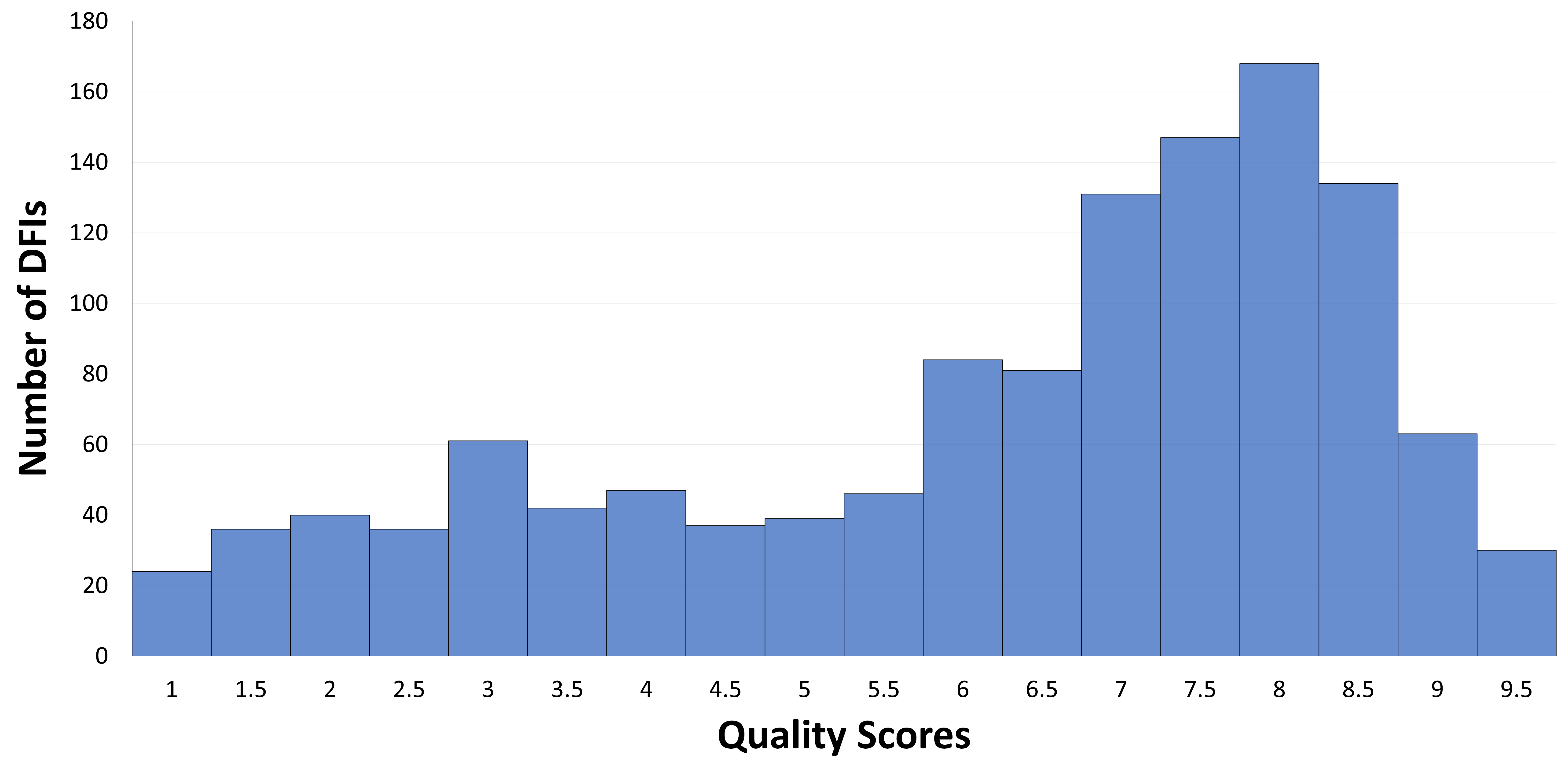}}
\caption{A histogram detailing the quality score distribution of the dataset annotated with the new quality scale and consisting of 1,245 DFIs.}
\label{quality_scores_dataset}
\end{figure}

The reference set was constructed by a glaucoma ophthalmologist with 30 years of experience (EZB). The reference set was open for viewing and comparison in the subsequent process of scoring individual DFIs. Two ophthalmologists, (EZB) and a senior resident (HP), provided independent quality annotations for an image set of 1,245 DFIs. Each DFI was independently scored on a scale of 1-10 by each of the two ophthalmologist. \textcolor{black}{When grading, the ophthalmologists referred to the elements of resolution, focus, contrast, brightness, artifacts, overall haze, uneven illumination and the ability to detect fine details, while the reference point for scoring was primarily the optic disc and the peri-papillary retina. That means that DFI artifacts which did not affect the area of interest, such as edge haze and thin lashes, were not considered during grading.} \textcolor{black}{ The two ophthalmologists discussed concomitantly each DFI and provided a joint score}. An example of the quality scale can be seen in Figure \ref{quality_scale_ref}. A breakdown of the distribution of the scores can be seen in Figure \ref{quality_scores_dataset}. 
To evaluate the quality of the DFIs, ophthalmologists particularly focused on the visibility of the optic disc and its surrounding area. A high grade was given if the details of this area were clearly visible with fine details, and the score was decreased if the visibility was compromised by factors such as over/under exposure, poor focus, haziness, or obscuration by eyelashes. The ophthalmologists did not consider the specific cause of any decreased quality, as this was not relevant to the clinical application of our scale. Instead, they compared the DFIs to a novel scale we constructed in order to easily determine the appropriate grade for each image.

It is important to note that blurriness in DFIs can be caused by various ocular pathologies, including corneal edema, cataracts, and vitreous opacity, as well as technical issues. However, these factors were not taken into account in the grading process, as the focus of our analysis was on the visibility of the optic disc and its surrounding area.

The new quality scale has several advantages compared to previous methods. First, it provides a more detailed assessment of image quality, with a higher resolution compared to other works. Second, since it was developed by ophthalmologists, it is more easily interpretable by them, making it more suited for clinical practice. Third, grading the images while having the scale open for scrutiny as part of grading each image greatly assists in scoring and prevents a “drift” in the scores secondary to the quality of the specific analyzed database. Fourth, the quality score allows for greater flexibility in determining the threshold for discarding images or considering its impact on the diagnosis process. Fifth, the scale was created using public databases, enabling other researchers to use it in their own work and compare their results.

\begin{figure*}[ht]
\centerline{\includegraphics[width=\textwidth]{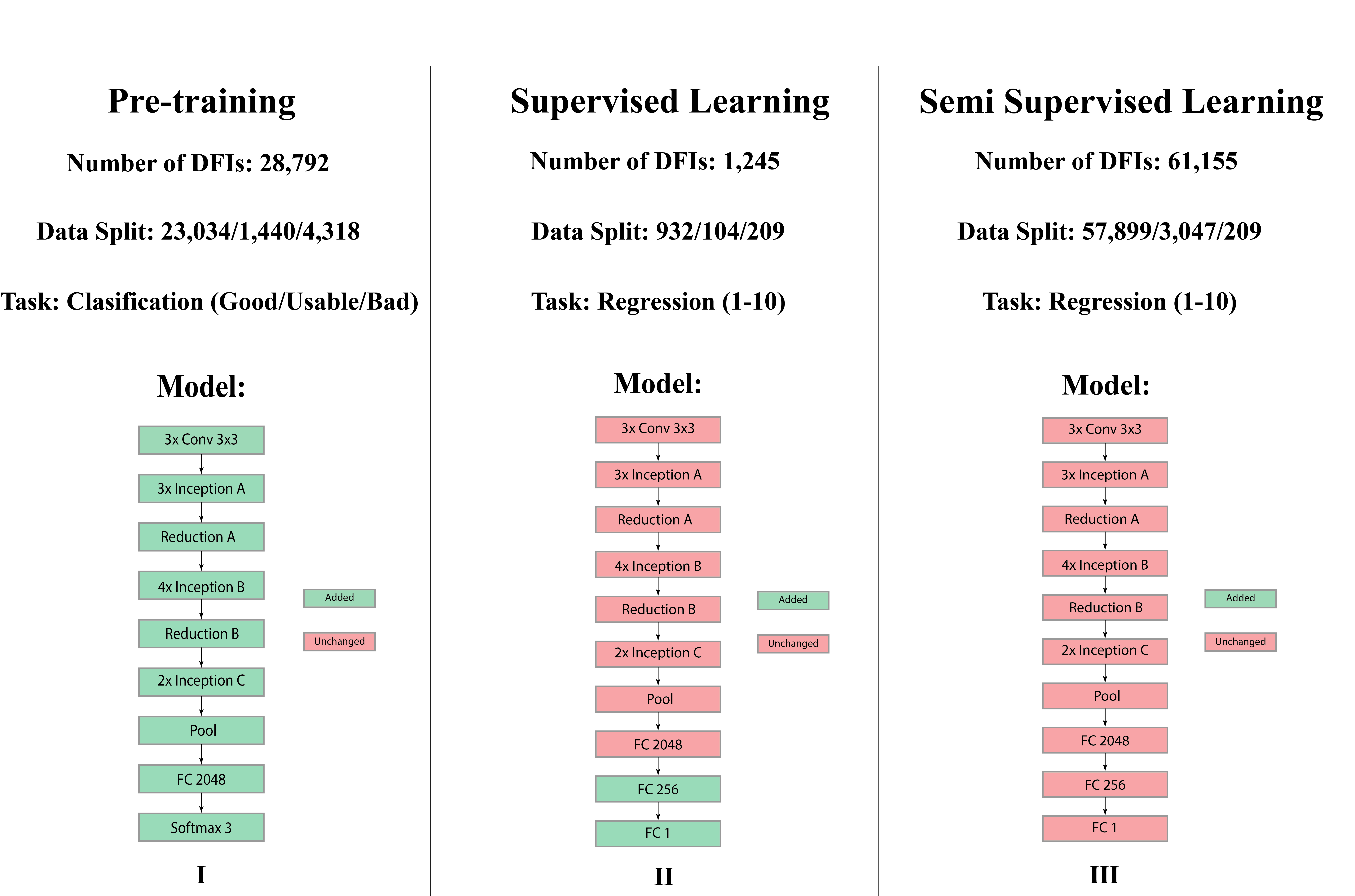}}
\caption{FundusQ-Net model development. The model is first pretrained using the EyeQ dataset and for a classification task (step I). Then the pretrained model is further trained for the quality grading regression task using the newly annotated dataset (step II). Finally, the semi-supervised learning is used to boost the model performance by leveraging unlabeled data (step III). The final model, i.e. following step I-III, is denoted FundusQ-Net.}
\label{pipeline}
\end{figure*}

Table \ref{tab:databases} summarizes all the databases used in this research. The quality scores distribution for the 1,245 DFIs can be seen in Figure \ref{quality_scores_dataset}.

\subsection{Databases}

\subsubsection{Quality scale definition and supervised learning}
The REFUGE database contains 1200 DFIs of 600 Chinese subjects captured for the purpose of glaucoma diagnosis \cite{Orlando2020REFUGEPhotographs}. The acquisition process involved the utilization of two different cameras: the Zeiss Visucam 500 and the Canon CR-2. Of the total DFIs, 400 were captured using the Zeiss camera, while the remaining 800 DFIs were captured using the Canon camera. For the purpose of the scale, 84 DFIs captured by the Zeiss camera were randomly chosen, since it has a higher quality than the Canon fundus camera.

The ORIGA database contains 650 DFIs of subjects of Malay Singaporean origin, collected by the Singapore Eye Research Institute for the purpose of glaucoma analysis and research \cite{Zhang2010ORIGA-lightResearch}. The subjects’ ages range from 40-79, and 25.8\% of the images are of glaucomatous patients, with images of males comprising 52.0\% of the total database. A total of 145 DFIs from the ORIGA database were randomly chosen for quality annotation using the new fundus quality scale.

The Drishti-GS database is a glaucoma-focused database, comprised of 101 DFIs collected consensually from visitors to the Aravind eye hospital in Madurai, India. The glaucomatous patients were chosen by clinical investigators during examinations and the healthy subjects were chosen from people undergoing routine refraction test. The subjects ranged in age from 40-80 years, with a male/female distribution of approximately 50\% \cite{Sivaswamy2014Drishti-GS:Segmentation}. A total of 21 DFIs from the Drishti-GS database were randomly chosen for quality annotation using the new fundus quality scale.

The LEUVEN database (IRB No. S63649) is a new, private database which contains 37,345 DFIs from 9,965 unique patients. There are 874 unique labels in the database, of which 61 describe variants of glaucoma. The labels include pathology diagnoses such as glaucoma or myopia, as well as procedures such as trabeculectomy or LASIK \cite{Hemelings2020AccurateLearning}. A total of 995 DFIs were randomly chosen.

\begin{table*}[hb!]
\centering
\begin{threeparttable}
\begin{tabular}{@{}ccccccc@{}}
\toprule
Model                                              & Mean Absolute Error (MAE) & Root Mean Squared Error (RMSE) & Min. Error       & Max. Error &  95\% CI & p*                \\ \midrule
Model 1              & 0.77           &0.99           & \textless{}0.01            & 4.04       & {[}0.69, 0.85{]}            & -                \\ \midrule
Model 2                  & 0.66         &0.89             & 0.01            & 4.12       & {[}0.58, 0.74{]}            & \textless{}0.01 \\ \midrule
\textbf{Model 3 - FundusQ-Net} & \textbf{0.61}          &\textbf{0.81}            & \textbf{\textless{}0.01} & \textbf{3.70}       & \textbf{{[}0.54, 0.68{]}}            & \textless{}0.01 \\ \bottomrule
\end{tabular}
\begin{tablenotes}
   \item[*] The p value is calculated by applying the Wilcoxon-signed-rank test between the results of the previous model and the current results. 
  \end{tablenotes}
\caption{Results on the internal test set (n=209). Model 1 was pre-trained on ImageNet \cite{Russakovsky2015ImageNetChallenge}. Model 2 was pre-trained on the EyeQ dataset \cite{Fu2019EvaluationColor-spaces}. Model 3 was pre-trained on the EyeQ dataset and was trained using pseudo-labeled DFIs \cite{Fu2019EvaluationColor-spaces}. }
\label{tab:internal-test-results}
\end{threeparttable}
\end{table*}

In determining the number of images to include in our analysis, we aimed to randomly sample approximately 20\% of the images from the DRISHTI, ORIGA, and REFUGE databases. For the LEUVEN database, which was particularly large, we limited our sample to 995 images, which were selected randomly. This sampling approach was chosen in order to ensure that our results were representative of the overall population of images contained in each database, while also taking into consideration the size of the datasets.

\subsubsection{Semi-supervised learning databases}
The EyePACS database is a diabetic retinopathy database, containing 88,702 DFIs. Images in this database were captured by a variety of models and types of cameras \cite{Fu2019EvaluationColor-spaces}. 59,910 DFIs from this database were used for semi-supervised learning. 

\subsubsection{Pre-training databases}
A total of 28,792 DFIs from the EyePACS database were annotated for quality by Fu et al. \cite{Fu2019EvaluationColor-spaces}, using the labels ``Good", ``Usable" and ``Reject". These images are known as the EyeQ database. In our research, we used the EyeQ database for pre-training.

\subsubsection{External test set}
The DRIMDB database contains 216 DFIs with three classes: ``Good" (n=125), ``Poor" (n=69) and ``Outlier" (n=22) \cite{Sevik2014IdentificationMethods}. The ``Good" and ``Poor" images are images of the fundus, whereas the outliers are images of the external eye or random objects. Due to difficulty translating our grading scale to the outlier category, only the 194 DFIs from the ``Good" and ``Poor" classes were used in this work, for the purpose of evaluating the generalization of the DL model to an external database.

\subsection{Quality assessment neural network}
We propose a DL model which is based on the Inception-V3 architecture. The model was developed using a total of 89,947 images from 6
databases, of which 1,245 were labeled by the specialists using the new scale system, and the remaining 88,702 images were used for pre-training and semi-supervised learning using a pseudo-labeling approach. All of the experiments were performed using a Dell 740XD cluster with 3 GTX Quadro RTX 6000 and 512 GB RAM Cards.
Figure \ref{pipeline} illustrates the development steps of the FundusQ-Net model. Step I involves pre-training a deep learning (DL) model using the EyeQ dataset for the multi-class classification task (good/usable/bad). Step II consists in using the pre-trained network from step I for the quality regression task using our new scale and newly annotated dataset. In step III, semi-supervised learning is used to improve the performance of the regression model trained in step II. For that purpose, pseudo-labeled DFIs from the EyeQ dataset were used.

\subsubsection{Data preprocessing}
In the preprocessing step, the DFIs were cropped to remove their black borders, in order to eliminate their influence on the neural network while still keeping the DFIs square. This was achieved by finding and removing the black outer rows and columns. The DFIs were subsequently resized to 224 x 224 pixels, to match the input layer of the Inception-v3 architecture \cite{Szegedy2016RethinkingVision}. 
Image augmentation was avoided, so that the quality of the DFI would not be accidentally modified.

\subsubsection{Pre-training}
Pre-training is a machine learning technique that involves training a model on a large dataset for a task that is not the task at hand. In a second step, the pretrained model is fine-tuned for the specific task of interest \cite{Devlin2018BERT:Understanding}. Pre-training can be particularly useful in cases where there is limited labeled training data available. It enables the model to learn patterns from the larger dataset that may also be useful for the task of interest  \cite{Kataoka2022Pre-TrainingImages}. Pre-training has been shown to improve performance in computer vision tasks \cite{Zhou2020ComparingRepresentations}.
In our work, we pretrained our model on the EyeQ database for a multiclass classification task (good, usable or bad quality).
Due to the limited number of quality graded DFIs using the new scale, pre-training can be used to help the DL model learn from the small database. Typically, DL models used for vision tasks are pre-trained on the ImageNet database, which contains millions of images organized into over 20,000 categories \cite{Russakovsky2015ImageNetChallenge}. However, due to the difference between natural images and DFIs, we performed two experiments: (1) pre-training on ImageNet and (2) pre-training on the EyeQ database. 
First, we compared several DL architectures, including some that were used in previous research interested in developing DFI quality scoring algorithms. This included AlexNet, used by Saha et al. \cite{SK2018AutomatedTelemedicine}, and DenseNet, used by Fu et al \cite{Fu2019EvaluationColor-spaces}. We also evaluated Inception-V3 and Xception \cite{Szegedy2016RethinkingVision} \cite{Chollet2016Xception:Convolutions}. The dataset used for this comparison was the EyeQ dataset. We selected the Inception-V3 architecture which yielded the highest performance for the pre-training task. The confusion matrices for each architectures evaluated are presented in supplement A. The loss function used during this phase was categorical cross-entropy, and the the optimizer used was Adam \cite{Gordon-Rodriguez2020UsesLearning} \cite{Kingma2014Adam:Optimization}.

\subsubsection{Supervised learning}
Following the previous step, the model was modified to perform the regression task. This was done by replacing the classifier layer with a new fully-connected layer, which was connected single neuron at the end. The model with the transferred weights, was then trained and evaluated using the 1,245 DFIs labeled with the new quality scale and with a train/validation/test split of 932/104/209. This is represented by stage II in Figure \ref{pipeline}. The database was stratified for the training, validation and test sets according to the quality scores, to guarantee an equal representation of each quality class. Following the pre-training step, the loss was changed to root mean squared error for the regression task (Fig. \ref{pipeline}, step II and III) \cite{Hodson2022Root-mean-squareNot}. The optimizer remained Adam \cite{Kingma2014Adam:Optimization}.

\subsubsection{Semi-supervised learning}

\begin{figure}
\centerline{\includegraphics[width=\columnwidth]{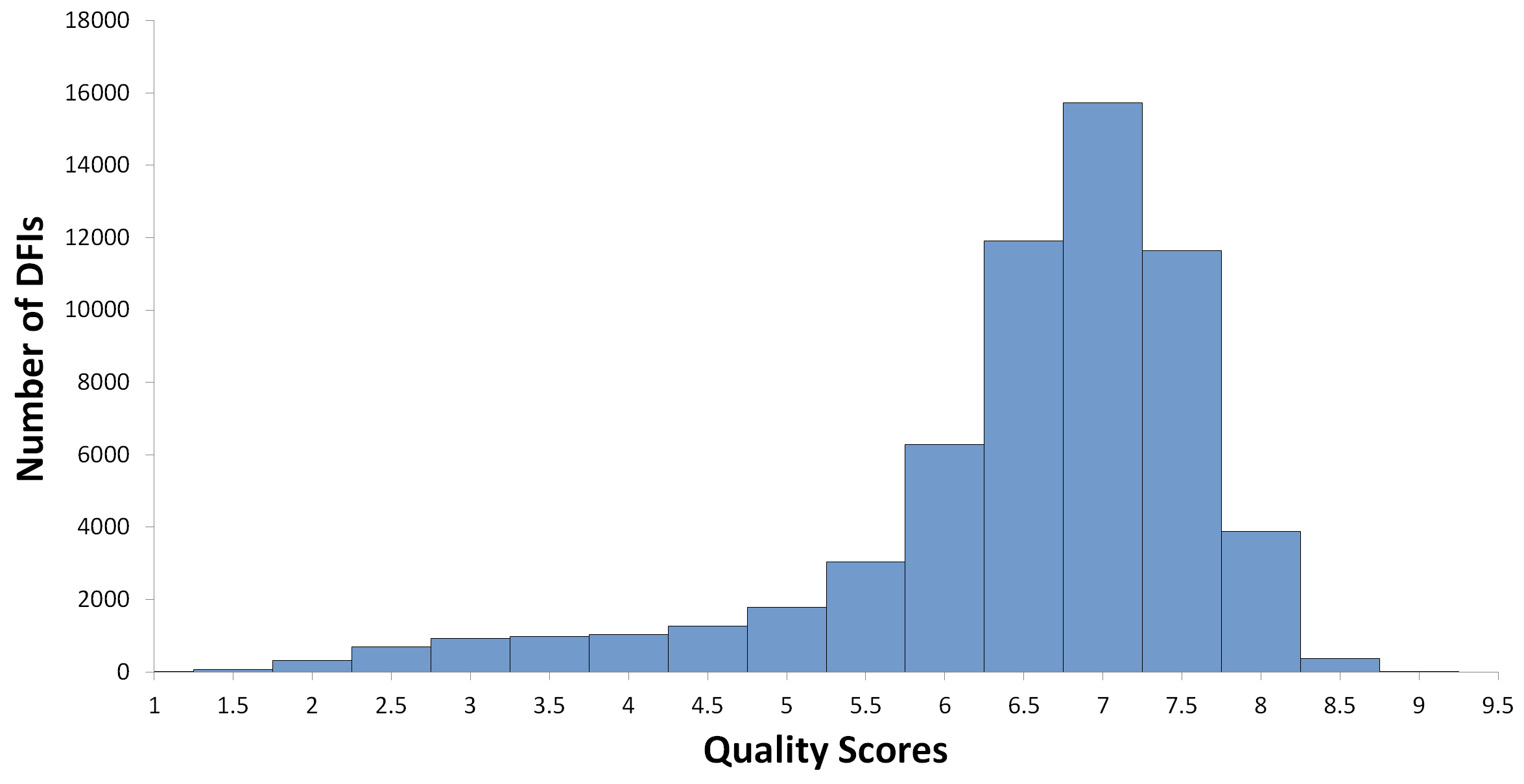}}
\caption{A histogram detailing the quality score distribution of the pseudo-labeled images from the EyeQ database.}
\label{pseudolabels_histogram}
\end{figure}

In supervised learning, classifiers require labeled data to train. However, creating these labels is often challenging, since they involve a time consuming annotation process by expert human annotators. Unlike traditional supervised learning, semi-supervised learning enables the use of large amounts of unlabeled data together with small amounts of labeled data to boost performance \cite{Zhu2005Semi-SupervisedSurvey}. In this work we used the pseudo-labeling \cite{Lee2013Pseudo-LabelNetworks} approach to semi- supervised learning, due to the limited number of graded DFIs available. Two models were utilized using this approach: a teacher and a student. The teacher was responsible to generate labels for unlabeled data, which are called ``pseudo-labels". The student utilized the unlabeled data with their pseudo-labels in addition to the labeled data when training, eventually surpassing the performance of the teacher.  \cite{Pham2020MetaLabels}.  The EyePACS database was chosen as a source of unlabeled DFIs \cite{Fu2019EvaluationColor-spaces}. The DFIs that are also included in EyeQ were excluded, resulting in 59,910 unlabeled DFIs that were used for the pseudo-labeling step.
After creating model II, which is featured in Figure \ref{pipeline}, it was used to create pseudo-labels for the EyePACS database. Following this step, the student model was trained using the labeled and pseudo-labeled data.
The train/validation/test split was 57,899/3,047/209. A histogram of the pseudo-labels is available at Figure  \ref{pseudolabels_histogram}.

\subsection{External Validation}
After training the Inception-V3 model, the DRIMDB database was used to perform external validation \cite{Sevik2014IdentificationMethods}. The final DL model inferred quality scores for each DFI from DRIMDB, which was then translated into ``Good" and ``Poor" according to the threshold 6.5 that was recommended by our consulting ophthalmologists. The average, standard deviation (STD) and maximal and minimal score were reported for each class, as well as the overall accuracy, sensitivity, specificity, Matthhew's Correlation Coefficient (MCC) and the Area Under the Curve (AUC) \cite{Chicco2020TheEvaluation} \cite{Brahmachari2013AreaCurve}. The model was benchmarked against different state-of-the-art works in order to show the value of using machine learning in this problem.

\subsection{Performance measures}
Since the network solves a regression problem, rather than a classification problem, the metrics chosen were Mean Absolute Error (MAE), Standard Deviation of the errors and the maximal error. To measure the improvement made by each step, the Wilcoxon-signed-rank test was used \cite{Rey2011Wilcoxon-Signed-RankTest}. 

\begin{figure*}[!t]
\centering
\begin{adjustbox}{scale=0.7}
\centerline{\includegraphics[width=\textwidth]{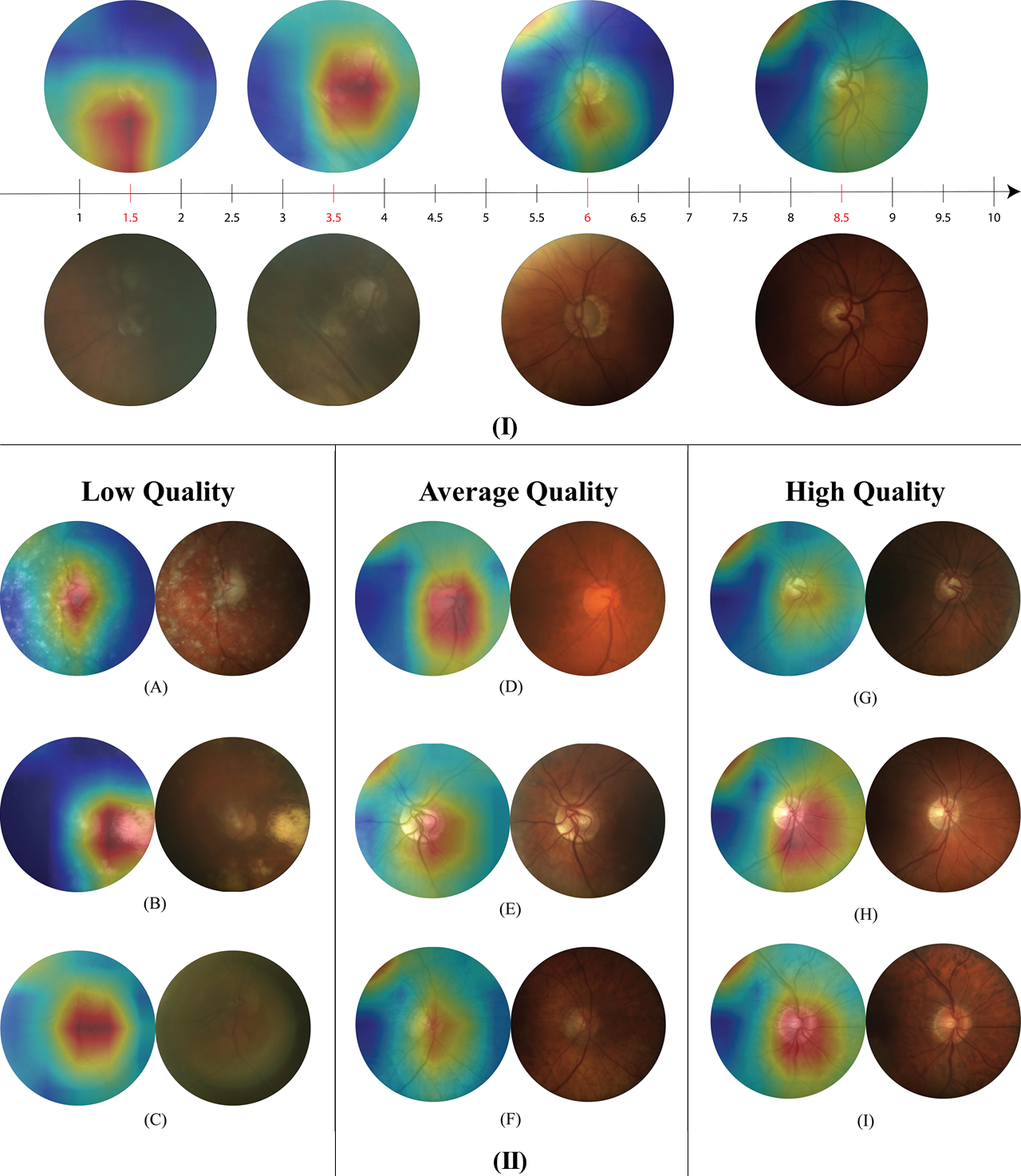}}
\end{adjustbox}
\caption{\textcolor{black}{Gradient-weighted Class Activation Map (Grad-CAM) analysis of FundusQ-Net for DFI with different quality scores. The CAM heatmaps are overlaid on the corresponding DFI. These heatmaps highlight the regions of the image that the model focuses on when making predictions. (I) a set of DFI examples are superimposed on the quality scale. (II) Nine additional DFI examples. DFIs A, B and C have a relatively low-quality score (2.5, 3 and 3 respectively), DFIs D, E and F have an average quality score 
(5, 6.5 and 6.5 respectively). DFIs H, G have a high-quality score (8, 8 and 9 respectively).} }
\label{cam_scale}
\end{figure*}

\begin{figure}[t]
\centerline{\includegraphics[width=\columnwidth]{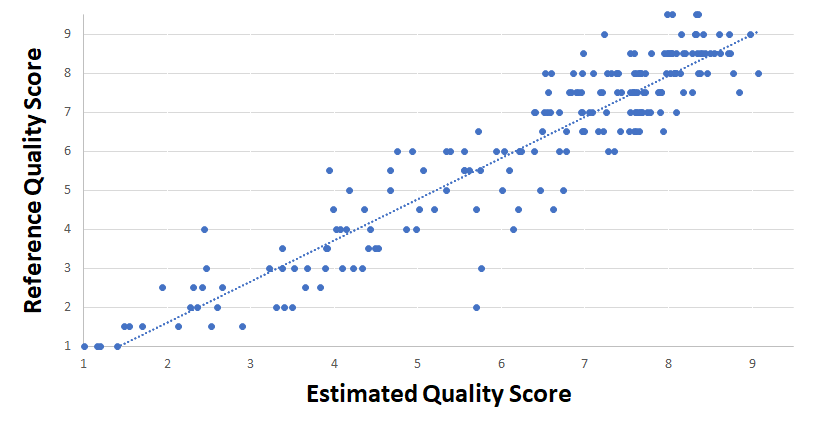}}
\caption{A linear fit between the DL inferred quality score (x) and the reference quality score (y) when tested on the internal test set (n=209). The achieved $R^2$ is 0.87.}
\label{linearfit_scores}
\end{figure}

\begin{figure}[!t]
\centerline{\includegraphics[width=\columnwidth]{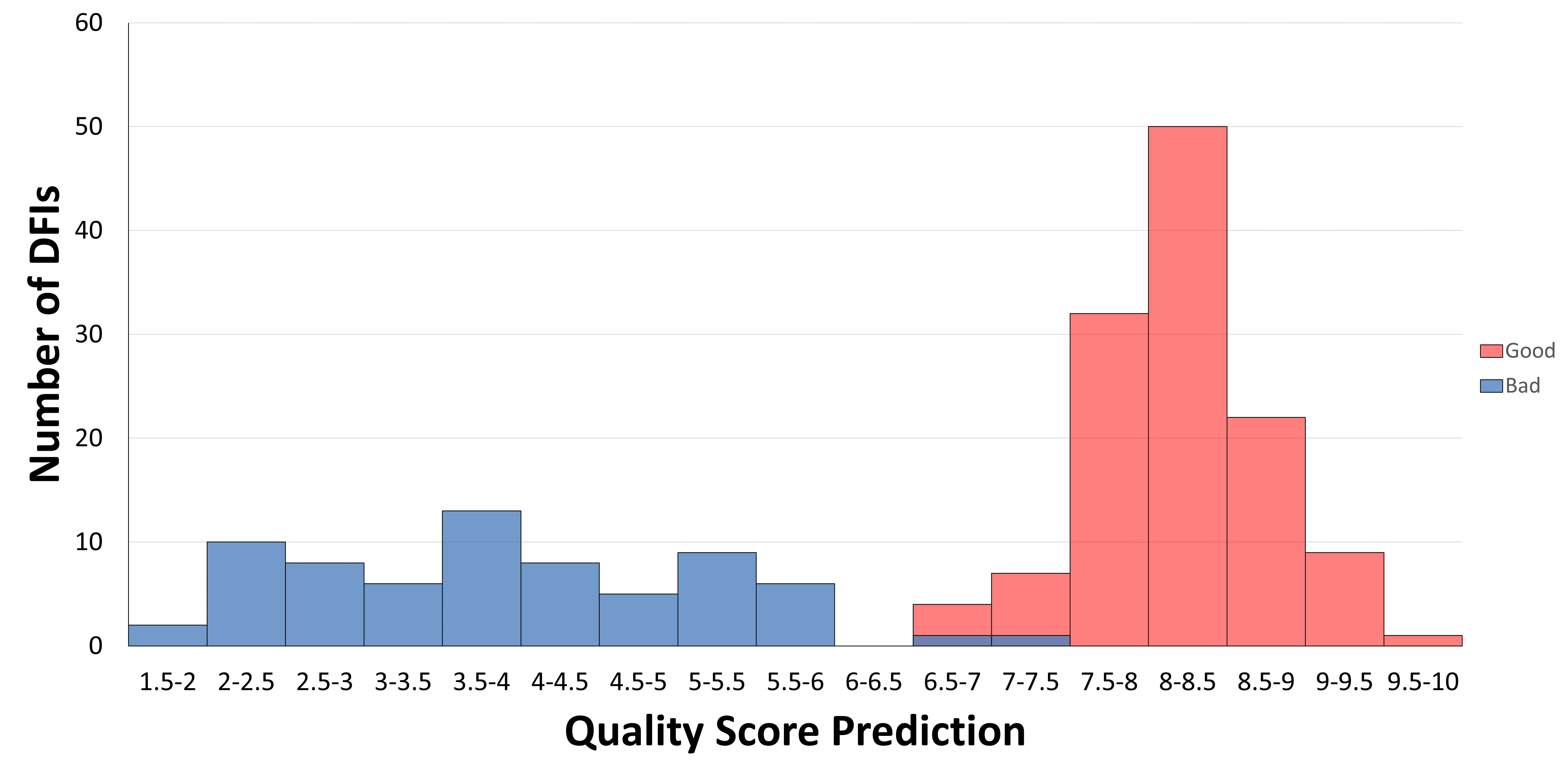}}
\caption{A histogram detailing the quality score estimation of the low and high quality DFIs on the external DRIMDB test database (n=194).}
\label{drimdb_joint_histogram}
\end{figure}

\begin{figure*}[ht]
\centerline{\includegraphics[width=\textwidth]{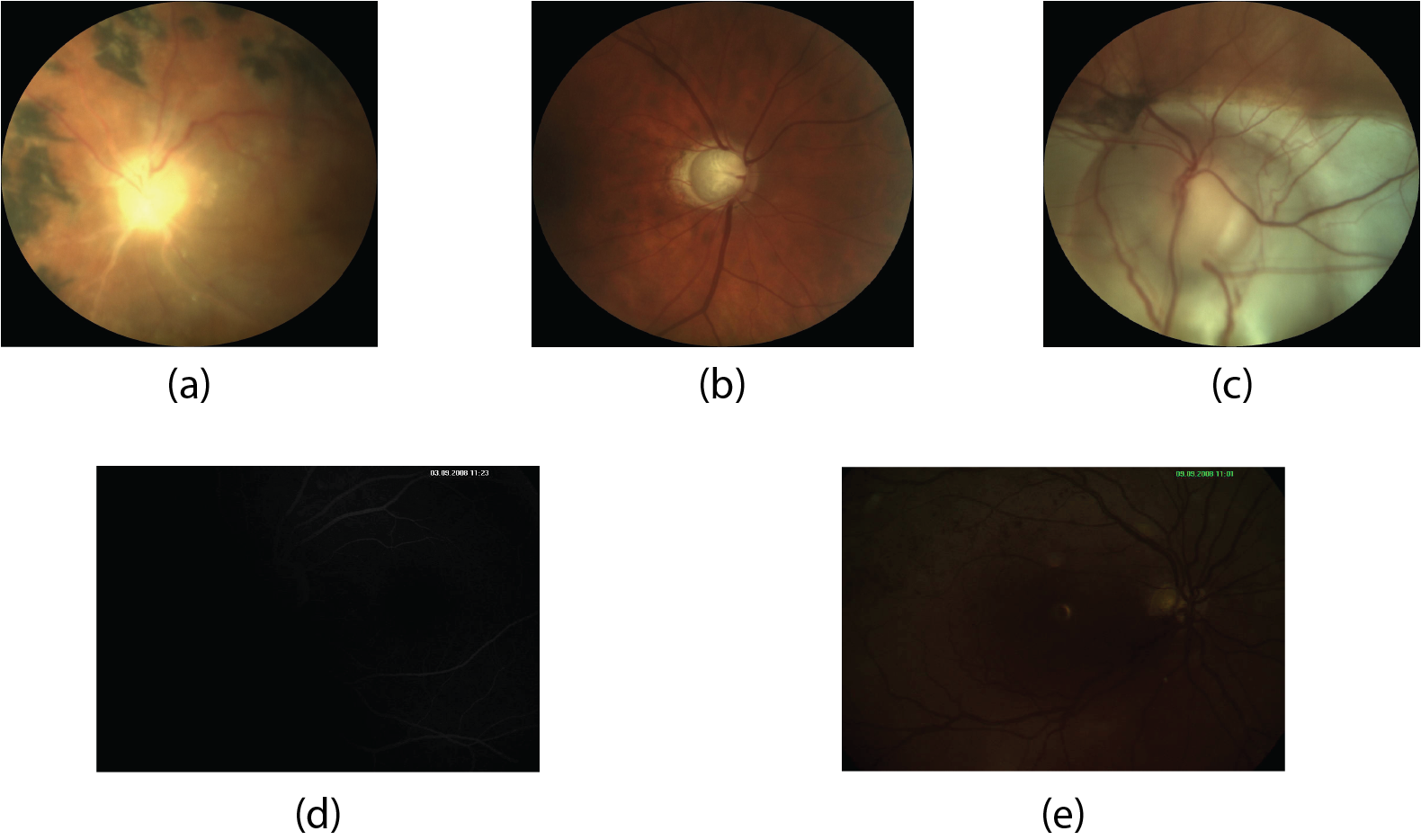}}
\caption{A figure displaying 5 selected outliers from both internal and external test sets. (a): a DFI where the quality of the retina is better than the disc, (b): a DFI where the quality of the disc is better than the quality of the retina, (c): A DFI of a very pathological eye, (d) + (e): the two outliers from the external test set}
\label{outliers}
\end{figure*}

\section{Results}

\subsection{Internal test set}
The results of the pre-training experiment on the internal test set are summarized in Table \ref{tab:internal-test-results}. Model 1, which was pre-trained using the ImageNet database achieved an MAE of 0.77$\pm$0.08 and a Root Mean Square Error (RMSE) of 0.99. The maximal error was 4.04 and the minimal error was \textless{}0.01. Model 2, which was pre-trained on the EyeQ database, achieved an MAE of 0.66$\pm$0.08 and an RMSE of 0.89. The maximal error was 4.12 and a minimal error was 0.01.

As expected, performing pre-training using DFIs yield lower MAE when compared to using ImageNet. Model 3 (FundusQ-Net) achieved an MAE of 0.61$\pm$0.07, and an RMSE of 0.81. The maximal error was 3.7 and the minimal error was \textless{}0.01 (Table \ref{tab:internal-test-results}). This means that the pseudo-label method was able to reduce the mean error by 7.6\%, the maximal error by 10\% and the STD by 14.5\%. Applying the Wilcoxon-signed-rank test between two subsequent models yields a p-value smaller than 0.01, demonstrating that each model has achieved meaningful improvement compared to the previous one. The linear fit shown in Figure  \ref{linearfit_scores} achieves an $R^2$ score of 0.87, suggesting a very good fit between our estimated scores and the reference quality scores. This reflects that the DL model can accurately determine the quality score of the DFI.

\subsection{External test set}
The histograms of the quality scores for the DRIMDB database can be seen in Figure \ref{drimdb_joint_histogram}, and the table containing the result of the experiment can be seen in Table \ref{tab:drimdb-results}.
Using a threshold of 6.5, our model achieved an accuracy of 99\%, with a sensitivity of 98.4\% and specificity of 100\%. In addition, the model has an of MCC 0.98 and an AUC score of 0.999  \cite{Chicco2020TheEvaluation}. This suggests that 6.5 is the correct threshold to use for this specific database, and demonstrates the flexibility in our proposed quality scale. In addition, our model outperforms performance reported in previous works when comparing accuracy and specificity, while being a very close second place when comparing the sensitivity. To determine the appropriate threshold value, input was sought from the expert ophthalmologists. While remaining blind to the classification results of FundusQ-Net on the external dataset, they set 6.5 as being a meaningful threshold for distinguishing between ``good" and ``bad" quality DFIs.

\begin{table*}[b!]
\centering
\begin{threeparttable}
\begin{adjustbox}{max width=\textwidth}
\begin{tabular}{@{}ccccccccc@{}}
\toprule
Method & \textcolor{black}{Quality Estimation Task} & \textcolor{black}{Overall Dataset Size} & \textcolor{black}{\begin{tabular}[c]{@{}c@{}}Semi-supervised\\ Learning Used\end{tabular}} & \textcolor{black}{Explainability} & Sensitivity & Specificity & Accuracy & AUC \\ \midrule
Shao et al. \cite{Shao2017AutomatedStructure} & \textcolor{black}{\begin{tabular}[c]{@{}c@{}}Binary\\ (Accept/Reject)\end{tabular}} & \textcolor{black}{6,314} & \textcolor{black}{No} & \textcolor{black}{None} & 94.1\% & 89.3\% & 89.1\% & 0.863 \\ \midrule
Chalakkal et al. \cite{Chalakkal2019QualityLearning} & \textcolor{black}{\begin{tabular}[c]{@{}c@{}}Binary \\ (Good/Bad)\end{tabular}} & \textcolor{black}{8,507} & \textcolor{black}{No} & \textcolor{black}{None} & 97.6\% & 97.8\% & 97.7\% & - \\ \midrule
Karlsson et al. \cite{Karlsson2021AutomaticScale} & \textcolor{black}{\begin{tabular}[c]{@{}c@{}}Regression\\ (0.0-1.0 Scale)\end{tabular}} & \textcolor{black}{1,232} & \textcolor{black}{No} & \textcolor{black}{None} & 96.9\% & 92.7\% & 95.4\% & 0.992 \\ \midrule
\textbf{FundusQ-Net} & \textcolor{black}{\begin{tabular}[c]{@{}c@{}}Regression\\ (EZB Quality Scale, 1-10)\end{tabular}} & \textcolor{black}{\textbf{89,947}} & \textcolor{black}{\textbf{Yes}} & \textcolor{black}{\textbf{CAM Analysis}} & \textbf{98.4\%} & \textbf{100\%} & \textbf{99.0\%} & \textbf{0.999} \\ \bottomrule
\end{tabular}
\end{adjustbox}
\caption{\textcolor{black}{Comparison of FundusQ-Net with State-of-the-Art Machine Learning Models for DFI quality estimation and performance on DRIMDB.}}
\label{tab:drimdb-results}
\end{threeparttable}
\end{table*}

\section{Discussion and conclusion}
In this paper we sought to establish a new, meaningful quality scale for DFIs. We further developed a DL algorithm for quality grading of DFI. For that purpose we combined the state-of-the-art Inception-V3 DL architecture with field specific pre-training and pseudo-labeling. Overall, high performance was obtained by the final model, denoted FundusQ-Net, with a MAE and confidence interval of 0.61 (0.55-0.69) on the test set and a generalization accuracy of 99\% on the external DRIMDB. Furthermore, we demonstrated that domain pre-training and pseudo-labeling improved the model performance significantly from MAE of 0.76 to 0.61 (p\textless0.05, see Table \ref{tab:internal-test-results}).

There are multiple motivations for performing a regression task against a quality scale versus a binary classification task (good/bad). These include: (1) Using a binary classification requires setting an arbitrary cutoff point to differentiate between good and bad quality images. This can be subjective and influenced by the grader's experience and beliefs. In contrast, a multi-step quality scale allows us to determine a less biased and more objective cutoff for sufficient quality; (2) Different clinical applications may require different quality thresholds. For example, the diagnostic criteria for glaucoma might require a higher quality score than for geographic atrophy, and vice versa. With a quality scale, we can adjust the cutoff to suit the specific clinical question at hand; (3) The quality threshold may also vary depending on the clinical setting and the availability of resources. In a resource-constrained environment or with a fully automated system, the pictures may be taken at a lower resolution by less qualified staff, resulting in lower quality images on average. In this case, using a lower quality threshold would prevent unnecessary repeat examinations and reduce the burden on specialist clinics, as well as avoid long distance travel for patients.

We performed Class Activation Map (CAM) analysis using Grad-CAM to examine the decision-making process of our CNN model, FundusQ-Net, for the task of assessing the quality of fundus images \cite{Zhou2015LearningLocalization} \cite{Selvaraju2017Grad-CAM:Localization}. Several fundus images of different qualities were considered in the CAM analysis. 
The results of the CAM analysis, shown in Figure \ref{cam_scale}, indicate that the model primarily focuses on two mains areas, namely: the optic disc region and some more distant areas that contained vasculature. The primary focus on the optic disc region in low quality DFI is meaningful since the clarity in visualizing the optic disc and its close surrounding was a primary criterion in the grading protocol. In the higher quality DFIs, the attention of the network to more distant areas containing vasculature is also sensible since in the case of DFIs with a good visualization of the disc then the annotators focused on the more distant surrounding in deciding for the exact grade. Overall, the CAM analysis results provide insights into the decision-making process of FundusQ-Net and help to better understand the model's strengths and limitations for the task of assessing the quality of fundus images. By examining the regions of the image that the model focuses on when making predictions, we were able to identify key features and patterns related to image quality that the model is learning to identify. These findings can inform future work on improving the performance of FundusQ-Net.

\subsection*{Error Analysis}
\textcolor{black}{Error analysis was conducted to gain a better understanding of the cases in which poor performance was observed for FundusQ-Net. Specialists were consulted to define outliers in the test set as DFIs with an absolute difference of more than 1.5 between their actual and estimated scores. Upon analyzing 11 of the outliers, it was found that the estimated score was higher than the actual score in 54.5\% of cases (n=6), increasing to 80\% (n=4) when the 5 largest differences were considered.}

\textcolor{black}{Several factors may contribute to these errors. One possibility is the discrepancies between the quality of the optic disc and the quality of the retina, which was identified in 36\% (n=4) of cases and found to be of higher quality in the optic disc area in 75\% (n=3) of cases. Examples of this can be seen in DFIs (a) and (b) in Figure \ref{outliers}. Another factor that may have played a role is a very pathological eye, as illustrated in example (c) in Figure \ref{outliers}. This could have caused the DFI to be very different from most training set examples, potentially leading to mistakes in prediction.}

\textcolor{black}{For the external test set, the threshold was set at 6.5. Two misclassified examples were found, both from the ``Bad" class and misclassified as ``Good". Upon review, it is suspected that these errors may have been caused by the extreme darkness of these DFIs, for which there were no examples in the training set. To address this issue, it is suggested that the training set be augmented with low quality examples including highly contrasted, very pathological, and dark images. Both outliers are depicted in Figure \ref{outliers}.}
\subsection*{Limitations and future work}
The DRIMDB dataset was originally annotated using a different methodology than the new quality scale developed in our work, specifically a binary labeling protocol versus labeling over an ordinal scale. Despite this limitation, our model FundusQ-Net demonstrated high generalization performance on this dataset. In order to further validate FundusQ-Net we will need to annotate additional DFIs from external datasets using the newly developed scale. Another limitation of our study is the lack of public access to the source code for the benchmark models. This limits our ability to statistically benchmark FundusQ-Net against these models. 
We also plan to investigate the effects of different classification thresholds on the performance of FundusQ-Net and its potential applications in clinical practice, as the threshold for classifying an image as high or low quality may vary depending on the intended use and context.
\textcolor{black}{In addition, methods to improve the quality of a DFI, such as the one created by Yoo et al \cite{Yoo2020CycleGAN-basedPhotography}, can be used concomitantly to our approach in an attempt to reduce the number of low quality DFIs that would be excluded from subsequent analysis.}
Finally, further evaluation of the model in a clinical setting is necessary to fully understand its potential for use in real-world scenarios. Accordingly, in order to strengthen the added value of FundusQ-Net for its clinical utility, we plan to demonstrate the performance of FundusQ-Net within the scope of a clinical task that involves the assessment of DFIs. This could include, for example, using FundusQ-Net as a preprocessing step for an AI system that is capable of diagnosing glaucoma or other ocular conditions.

\subsection*{Conclusion}
In conclusion, we created a novel and open-access quality scale for fundus images. This scale has not been previously developed and is intended to provide a standardized tool for assessing the quality of fundus images. We developed a novel deep learning model (FundusQ-Net) that leverages in-domain pre-training and semi-supervised learning for the regression task of DFI quality estimation. We demonstrated the high performance of this model on both a local and an external test set. Finally, we provided a thorough error analysis that includes class activation maps and investigates the model decision process.

\section*{Acknowledgment}
\textcolor{black}{The research was supported by a cloud computing grant from the Israel Council of Higher Education, administered by the Israel Data Science Initiative. This research was partially supported by Israel PBC-VATAT and by the Technion Center for Machine Learning and Intelligent Systems (MLIS).}

\section*{Authors contributions}
\textcolor{black}{O.A. performed the data analysis (data preparation, deep learning model  developments, explainability analysis), contributed to the methodology and to the writing of the original draft. E.B. and H.P. designed the quality annotation scale, graded the DFI and provided clinical guidance along the research. J.V.E. and I.S. contributed the LEUVEN database and data curation. I.O. and J.M. contributed to the methods and results interpretation. J.B. contributed to the conceptualization, methodology, supervision of O.A. and wrote the original draft. All authors reviewed the manuscript and provided extensive comments.}

\textcolor{black}{\printbibliography}
\clearpage
\twocolumn[
\begin{@twocolumnfalse}
\begin{appendices}
\section{Confusion Matrices}
\centering 
\begin{adjustbox}{scale=1.2}
\hspace*{-2cm} 
\centerline{\includegraphics[width=\textwidth]{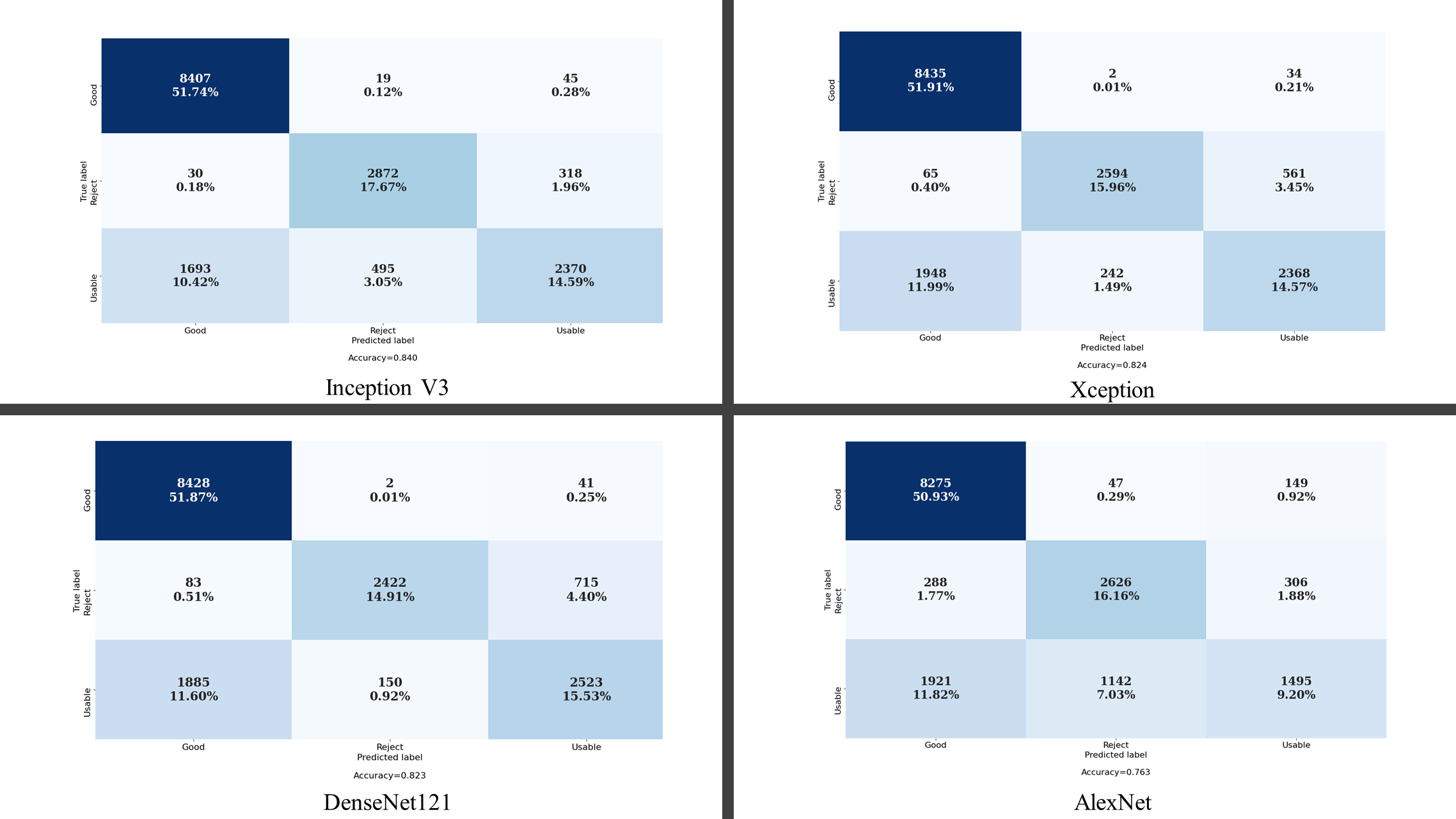}}
\end{adjustbox}
\captionof{figure}{The confusion matrices of the benchmarked models 
for the multiclass classification task (good/usable/bad) using the EyeQ dataset  \cite{Fu2019EvaluationColor-spaces}.}
\label{matrices}
\end{appendices}
\end{@twocolumnfalse}]
\end{document}